\documentclass{aa}  

\usepackage{graphicx}

\usepackage{siunitx}

\usepackage{txfonts}
\usepackage{todonotes}
\newcommand{\bb}[1]{\boldsymbol#1}

\begin{document}

   \title{Thermophysical model for icy cometary dust particles }

   \subtitle{}

% MM update
   \author{J. Markkanen\inst{1}
          \and
          J. Agarwal\inst{1,2}
          }
% MM update
   \institute{$^1$Max Planck Institute for Solar System Research, Justus-von-Liebig-Weg 3, 37077 G\"ottingen, Germany\\
   $^2$Institut f\"{u}r Geophysik und extraterrestrische Physik, Technische Universit\"{a}t Braunschweig, Mendelssohnstr. 3, 38106 Braunschweig, Germany\\
\email{markkanen@mps.mpg.de}}
 \date{}

% \abstract{}{}{}{}{} 
% 5 {} token are mandatory
 
  \abstract
  % context heading (optional)
  % {} leave it empty if necessary  
   {Cometary dust particles are subjected to various forces after being lifted off the nucleus. These forces define the dynamics of dust, trajectories, alignment, and fragmentation, which, in turn, have a significant effect on the particle distribution in the coma. }
  % aims heading (mandatory)
   {We develop a numerical thermophysical model that is applicable to icy cometary dust to study the forces attributed to the sublimation of ice.}
  % methods heading (mandatory)
   {We extended the recently introduced synoptic model for ice-free dust particles to ice-containing dust. We introduced an additional source term to the energy balance equation accounting for the heat of sublimation and condensation. We use the direct simulation Monte Carlo approach with the dusty gas model to solve the mass balance equation  and the energy balance equation simultaneously. }
  % results heading (mandatory)
   {The numerical tests show that the proposed method can be applied for dust particles covering the size range from tens of microns to centimeters with a moderate computational cost. We predict that for an assumed ice volume fraction of 0.05, particles with a radius, $r >> \SI{1}{\milli\metre,}$ at 1.35\,AU, may disintegrate into mm-sized fragments due to internal pressure build-up. Particles with $r < \SI{1}{\cm}$ lose their ice content within minutes. Hence, we expect that only particles with $r >$ 1cm may demonstrate sustained sublimation and the resulting outgassing forces.}
  % conclusions heading (optional), leave it empty if necessary 
   {}

 \keywords{Comets: general, Diffusion, Radiative transfer, Methods: numerical}

   \maketitle
%
%-------------------------------------------------------------------

\section{Introduction}
As comets approach the Sun, they start to eject gas and dust from their nuclei. The ejection of dust is mainly driven by the sublimation of volatile ices. Thermal modelling suggests that the sublimation of volatiles such as CO and CO$_2$ ices can lift large particles containing water ice off the surface, whereas sublimation of water ice can only lift smaller, ice-free particles \citep{Gundlach2020}. In addition, other mechanisms, such as the crystallisation of amorphous ices and impacts, may be capable of ejecting smaller icy particles. The ice-containing particles ejected from the surface may behave like mini-comets whose trajectories are affected by the recoil force due to anisotropic outgassing of water vapour. Indeed, particles whose trajectories cannot solely be explained by gravity, gas drag, and radiation pressure have been detected for the comets 103P/Hartley 2 \citep{Kelley2013, Kelley2015} and 67P/Churyumov-Gerasimenko \citep{Agarwal2016}. Furthermore, the sublimation of ices can build up pressure in the interior of a particle, which may lead to their disintegration while they are travelling from the inner to outer coma.

Understanding the outgassing of dust particles requires rigorous multiphysical modelling. Typical thermophysical models applied to the cometary nuclei, however, introduce various simplifications which make them inapplicable to small dust particles. More specifically, the models are often one dimensional, they assume wavelength-independent surface absorption and emission of electromagnetic radiation, the radiative heat transfer is treated as a one-dimensional transport by defining an effective heat transfer coefficient, and the gas flow is treated as the Knudsen flow \citep{Huebner2006, Prialnik2004, Gundlach2020}. For small particles, these approximations may introduce significant errors. The thermal evolution of small cometary dust particles is often modelled by assuming isothermal spherical particles and using the analytical Lorenz-Mie solution for the absorbed and emitted radiation and assuming that the sublimation only occurs on the surface of the particle \citep{LICHTENEGGER1991, Gicquel2012, Gicquel2016}. This is a valid assumption for small compact particles, but inside porous particles larger than tens of micrometers, temperature and pressure gradients may occur in a typical coma environment. Thus, a more realistic numerical model is needed to bridge the gap between small and very large particles.

We recently introduced a numerical method for analyzing the thermal properties of ice-free particles by treating the radiative heat transfer with the radiative transfer with the reciprocal transactions framework \citep{Markkanen2019}. Here, we extend the method for ice-containing particles by introducing an additional source term to the energy equation that accounts for the latent heat of sublimation and condensation. Because this term depends on the temperature and pressure, we solve the energy balance equation simultaneously with the mass balance equation. We apply the finite-element method (FEM) for the energy balance equation and the direct simulation Monte Carlo (DSMC) method with the dusty gas model for the mass transport. 

The article is organised as follows. In Section 2, we present the governing equations describing the physics of the problem. The numerical methods we used are explained in Section 3. Section 4 introduces the particle model. In Section 5, we present the validation of the developed method against the continuous gas flow model and apply the method to study thermal properties of icy cometary dust particles at 1.35\,AU. Finally, we present our conclusions in Section 6. 

\section{Governing equations}
Recently, we introduced a thermal model for the aggregated ice-free particles composed of submicrometer-sized dust grains in \cite{Markkanen2019}. Here, we will extend the method for dust particles containing ice inclusions. We assume that the dust and ice grains are intimately mixed in the microscopic scale. We also assume local equilibrium, in the case of which the solid and gas phases have the same temperature locally and the energy exchange between the solid and gas phases is neglected. 

\subsection{Energy balance}

The energy balance equation for the solid phase is written as
\begin{equation}
c_{\rm p} \rho_{\rm s} \frac{\partial T}{\partial t} - \nabla \cdot \kappa \nabla T = Q_{\rm r} + Q_{\rm l}
\label{eqenergy}
,\end{equation}
where $c_{\rm p}$ is the specific heat capacity, {\bf $\rho_{\rm s}$} is the density, $T$ is the temperature, and $\kappa$ is the conductive heat transfer coefficient. The right-hand side includes the volumetric source terms, the absorbed solar and thermally emitted and reabsorbed radiation $Q_{\rm r}$, and the latent heat of sublimation and condensation of ices $Q_{\rm l}$. The above values are macroscopic, that is, they are averaged over a small volume element larger than the microstructure of dust and ice grains. We also assume that the density and specific heat capacity of gas are many orders of magnitude lower than those of the solid phase and ignore the heat transport in the gas phase.

On the particle surface, we impose the Neumann boundary condition for the conductive flux as $\bb n \cdot \kappa \nabla T = 0$, where $\bb n$ is the outer unit normal vector. Thus, only the radiative heat flow can cross the boundary as the convective flow is neglected.

Energy related to the sublimation and condensation processes can be written as
\begin{equation}
Q_{\rm l} = q H
,\end{equation}
where $q$ is the sublimation rate and $H$ is the latent heat. By defining the saturation vapour pressure as \begin{equation}
P_{\rm sat} = A e^{-B/T}
\end{equation}
and using the experimentally obtained coefficients, $A$ and $B$ \citep{Fanale1984}, the heat of sublimation is obtained via the Clausius-Clapeyron relation and given by
\begin{equation}
    H = B \frac{R_{\rm g}}{\mu} 
,\end{equation}
in which $R_{\rm g}$ is the universal gas constant and $\mu$ is the molar mass of the gas.

The production rate is obtained via the Hertz-Knudsen formula as
\begin{equation}
    q = S(P_{\rm sat} - P) \sqrt{\frac{\mu}{2\pi R_{\rm g} T}}
    \label{eq_production_rate}
,\end{equation}
where $S$ is the surface-to-volume ratio in the microscale, and $P$ is the pressure in the micropores.

\subsection{Mass balance}
The mass balance equation for the gas and solid phases is written as
\begin{eqnarray}
\frac{\partial \rho_{\rm g}}{\partial t} + \nabla \cdot \bb J = q ,\\
\frac{\partial \rho_{\rm s}}{\partial t} = -q
,\end{eqnarray}
where $\rho_{\rm g}$ and $\rho_{\rm s}$ are the gas and solid phase densities, respectively, and $\bb J$ is the gas flux density. The macroscopic thermodynamic properties of gas such as the density, velocity, pressure, and flux emerge from the stochastic microscopic description of the state of gas given by the Boltzmann equation:\ 
\begin{equation}
    \frac{\partial f}{\partial t} + \bb v \cdot \nabla_{\bb r} f = C(f,f) + f_{\rm q}
    \label{eqboltzmann}
,\end{equation}
where $f(t,\bb r, \bb v)$ is the time ($t$) position ($\bb r)$ and velocity ($\bb v)$ dependent distribution density function, $f_{\rm q}$ is the source function, and $C(f,f)$ is the collision operator. Formally, the collision operator reads 
\begin{equation}
C(f,f) =   \int \int |\bb v - \bb v_*| (f_*'f' - f_*f) \sigma(\Omega) \,\text{d}\Omega\,\text{d}\bb v_*,
\end{equation}
where $\sigma$ is the kernel that describes scattering, the primed functions are the  post collision distribution densities, and the subscript * denotes the other particle in the collision pair.

The macroscopic number density $n$ can be computed by integrating the distribution density function over the velocity space as
\begin{equation}
    n(t, \bb r) = \int f(t,\bb r, \bb v) \,\text{d}\bb v
,\end{equation}
and the gas flux as
\begin{equation}
    \bb J(t, \bb r) = \int \bb v f(t,\bb r, \bb v) \,\text{d}\bb v.
,\end{equation}
We apply the DSMC method to solve Equation (\ref{eqboltzmann}), as described in Section 3.

\section{Numerical solution}

We discretise the domain of interest $\Omega$ with tetrahedral elements. Then we employ the FEM for the spatial dimension of the energy balance equation (\ref{eqenergy}) with the nodal testing $w^n$ and basis functions $u^m$. For the temporal dimension we will use the central finite difference formula 
\begin{equation}
\left. \frac{\partial T}{\partial t}\right \rvert_{t+\frac{1}{2}}  \approx \frac{T_{t+1} - T_{t}}{\tau} = L_{t+\frac{1}{2}}     
\end{equation}
and interpolate the other terms at $t+\frac{1}{2}$ as
\begin{equation}
L_{t+\frac{1}{2}} \approx \frac{L_{t+1} + L_{t}}{2}. 
\end{equation}
This results into the Crank-Nickolson scheme for the unknown coefficient vector $x_{t+1}$ given by
\begin{equation}
\label{eq_diff}
x_{t+1} = (M + \frac{\tau}{2} S)^{-1} (Mx_t + \frac{\tau}{2} Fx_t - \frac{\tau}{2}Sx_t + Fx_{t+1})    
,\end{equation}
where the mass and the stiffness matrices are defined as
\begin{equation}
M = \rho_{\rm s} c_{\rm p} \int_\Omega w^n u^m\,\text{d}V,    
\end{equation}
\begin{equation}
S = \int_\Omega \nabla w^n \cdot \kappa \nabla u^m\,\text{d}V,    
\end{equation}
and the force vector as
\begin{equation}
\label{eq_force}
F = \int_\Omega w^n (Q_{\rm r} + Q_{\rm l})\,\text{d}V.     
\end{equation}
Equation (\ref{eq_diff}) is strongly non-linear because of the force term. Hence, we use an iterative method with under-relaxation to solve the unknown coefficient vector, $x_{t+1}$.  
 
 We compute the absorbed solar and thermally emitted radiation,  $Q_r$, by using the radiative transfer with reciprocal transactions framework, as described by \cite{Markkanen2019}. Energy related to sublimation and condensation requires evaluation of the gas production rate, $q,$ which depends on the temperature, $T,$ and pressure, $P$. Thus, we need to solve the mass balance equation simultaneously with the energy balance equation. We apply the DSMC method, and its solution thus satisfies the Boltzmann equation, to solve the gas flow by assuming that the flow is a free molecular flow, that is, the mean free path of molecule-molecule collisions is much larger than the mean free path of molecule-grain collisions. The grains are assumed to be much heavier than the gas molecules and, thus, they are stationary, giving us the so-called dusty gas model. We consider the molecule-grain collisions stochastically, as in \cite{Ahmadian2019}.
 
 To compute the source function $f_{\rm q}$ in the DSMC, we introduce new (or remove the existing) DSMC molecules for each tetrahedral element with the index, $n,$ based on the production rate, $q^n$, at any given time step, $\Delta t_{\rm DSMC},$ with weight, $q^n \Delta t_{\rm DSMC}$. The velocity of the new DSMC molecule is drawn from the local solid-phase temperature-dependent Maxwell-Boltzmann distribution. Then the trajectories of the DSMC molecules are traced within a time step. To account for the collision operator, $C$, the molecule-grain scattering distance is drawn from the exponential distribution with the mean free path, $l_{\rm DSMC}$, and a new velocity for the gas molecule is drawn the Maxwell-Boltzmann distribution defined by the local temperature. Finally, the macroscopic thermodynamical properties, for instance, density, pressure, and flux are sampled by summing up the DSMC molecules in each tetrahedron and averaging them over the time frame. The process is repeated until the steady state or the energy equation time step is reached.
 
 The characteristic time scales for energy and mass transport can be very different. This means that we need to use different time steps ($\tau, \Delta t_{\rm DSMC}$) for the the energy and mass transport to get a sufficiently fast numerical method. The  DSMC time step, $\Delta t_{\rm DSMC}$, must be small enough such that the DSMC molecules cannot travel across a finite cell within a single time step, whereas $\tau$ is limited by the non-linearity and the convergence of the iterative method used to solve the energy balance equation. In practice, for cometary dust applications, the DSMC time step must be many orders of magnitudes smaller than the energy balance equation time step. Often, the gas flow reaches the steady state within the energy balance equation time step. In such a case, we stop the DSMC simulation and extrapolate the total gas production rate and the remaining ice mass at $t + \tau$. The extrapolation does not conserve ice and gas mass exactly when a tetrahedron runs out of ice within the time step $\tau$ as the ice mass cannot be negative. To avoid the problem, we run the DSMC again by setting the ice mass to zero for the tetrahedron that has run out of ice and calculate the production rates using the total mass loss at the given tetrahedron as
 \begin{equation}
 q_{t+1} =\frac{m_{t}}{\tau}  
 ,\end{equation}
 where $m_{t}$ is the ice mass at time $t$.

 \section{Particle model}
 
 We used a particle microstructure model derived by fitting the phase function of the coma of 67P/Churyumov-Gerasimenko in \cite{Markkanen2018_2}. The heat conduction coefficient was derived by fitting the superheating phase function of the same comet in \cite{Markkanen2019}.

The microstructure consists of submicrometer-sized organic monomers and micrometer-sized silicate monomers randomly deposited inside a particle with the porosity, $\Phi=0.6$. The effective heat capacity, $c_{\rm p} = \SI{750}{\joule \ kg^{-1}\kelvin^{-1} }$, and the effective density, $\rho_{\rm s} = \SI{1000}{\kg\metre^{-3}}$, are estimated from the porosity $\Phi$ and the silicate-to-organic volumetric ratio of 1/5. The size distributions of monomers follow a differential power law with the index of -3, and the minimum and maximum cutoff limits are $a_{\rm min} = \SI{65}{\nano\metre}$ and $a_{\rm max} = \SI{125}{\nano\metre}$ for the organic and $a_{\rm min} = \SI{0.6}{\micro\metre}$ and $a_{\rm max} = \SI{1.3}{\micro\metre}$ for the silicate monomers. The refractive index for the organic monomers is approximated to be the same as amorphous carbon and is taken from \citet{Jager1998}. For the silicate monomers the refractive index is taken from \citet{Dorschner1995}.  Here, we assume that ice is uniformly deposited in the micropores and the amount of ice is so low that it does not have an effect on the scattering properties. The microstructure model gives rise to the incoherent scattering properties of the volume elements which are used to evaluate the radiative part of the energy equation as described by \cite{Markkanen2019}. In addition, the gas mean free path, $l_{\rm DSMC}$, and the surface to volume ratio, $S,$ are derived from the microstructure model. The parameters used in the model are presented in Table \ref{KapSou}. 
 
\begin{figure}[htb]
%\begin{center}
\includegraphics[width=0.5\textwidth]{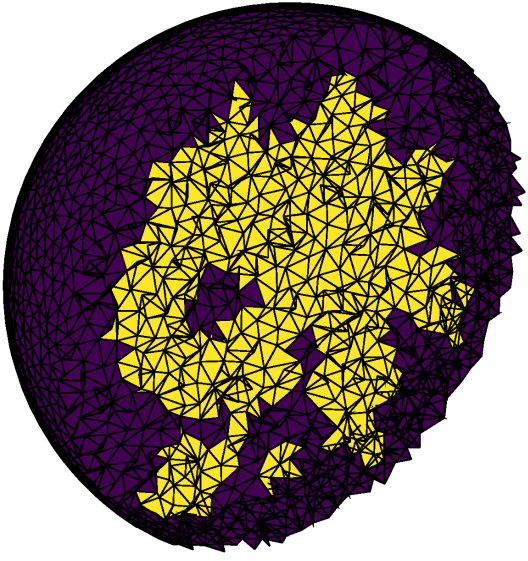}
\caption{Crosscut of an example computational mesh generated by the two-level Voronoi partitioning. The yellow tetrahedra correspond to the particle with a specified microstructure and the blue ones correspond to free space.}
%\end{center}
\label{fig0}
\end{figure}

 The macrostructure is described by the tetrahedral mesh in which each tetrahedron can be assigned to different microstructure, effective material properties $c_{\rm p}, \rho_{\rm s}, \kappa$, and ice content.  To generate the macrostructure model, we applied the hierarchical Voronoi partitioning algorithm with the parameters, $N_1 = 100$, $N_2=1000$, $P^{\rm rm}_1 = 1$, and $P^{\rm rm}_2 = 0.4,$ from \cite{Markkanen2015}. This creates particles with the macroporosity of 0.6 and the mean unit Voronoi cell size of one tenth of the particle size. Figure \ref{fig0} shows an example computational mesh generated by the two level hierarchical Voronoi algorithm. In the remainder of the paper, the size of the particle refers to the radius, $r,$ of the smallest sphere containing the particle if not defined otherwise.
 
 \begin{table}
      \caption[]{Simulation parameters.}
         \label{KapSou}
      $$
         \begin{array}{p{0.5\linewidth} p{0.1\linewidth} p{0.25\linewidth} ll}
            \hline
            \noalign{\smallskip}
            Parameter  & symbol  &  value \\
            \noalign{\smallskip}
            \hline
            \noalign{\smallskip}
            Density & $\rho_{\rm s}$ & \SI{1000}{\kg\metre^{-3}}     \\
            Specific heat capacity & c$_{\rm p}$ & \SI{750}{\joule \kg^{-1} \kelvin^{-1}}             \\
            Heat conduction coefficient & $\kappa$ & \SI{0.00025}{\watt\metre^{-1}\kelvin^{-1}}  \\
            Surface-to-volume ratio & $S$  & \SI{1.0e7}{\metre^{-1}}             \\
            Vapour pressure coefficient & $A$  & \SI{356.0e10}{\pascal}             \\
            Vapour pressure coefficient & $B$  & \SI{6141.667}{\kelvin}             \\
            Porosity & $\Phi$  & 0.6             \\
            Molecule-grain mean free path & $l_{\rm DSMC}$  & \SI{1.0e-6}{\metre}             \\
            \noalign{\smallskip}
            \hline
         \end{array}
     $$
   \end{table}

%--------------------------------------------------------------------
\section{Results}

\subsection{Validation}

To validate the DSMC implementation, we compared the results with the continuous gas flow model in porous media. The Fick's law states that the gas flux $\bb J$ is proportional to the gas density $\rho_{\rm g}$ gradient as 
\begin{equation}
    \bb J = -D \nabla \rho_{\rm g}
,\end{equation}
where $D$ is the diffusion coefficient. Here, the gas density $\rho_{\rm g}$ is defined as the gas mass per total volume including dust grains. Using the continuity equation, we have 
\begin{equation}
\frac{\partial \rho_{\rm g}}{\partial t} -\nabla \cdot D \nabla \rho_{\rm g} = S(P_{\rm sat}-P) \sqrt{\frac{\mu}{2\pi R_{\rm g}T}}.
\end{equation}
Writing the gas density in terms of pressure inside the micropores and using the ideal gas assumption,
\begin{equation}
    P = \frac{\rho_{\rm g} R_{\rm g} T} { \mu \Phi}
\end{equation}
leads to the following equation for the gas density
\begin{equation}
  \frac{\partial \rho_{\rm g}}{\partial t}- \nabla \cdot D \nabla \rho_{\rm g} = S \left(P_{\rm sat} - \frac{\rho_{\rm g} R_{\rm g} T} { \mu \Phi}\right) \sqrt{\frac{\mu}{2\pi R_{\rm g} T}}.
  \label{eq_gas}
\end{equation}

The diffusion coefficient for the random walk in three dimensions with the Maxwell-Boltzmann mean velocity $v_{\rm th}$ is given by
\begin{equation}
D = \frac{1}{6} l_{\rm DSMC}v_{\rm th} = \frac{1}{6} l_{\rm DSMC} \sqrt{\frac{8R_{\rm g}T}{\pi \mu}}.     
\end{equation}
Furthermore, by assuming the half Maxwell-Boltzmann distribution above the particle interface, giving the normal mean velocity of $v_{\rm n} = v_{\rm th}/4$ \citep{Huebner2000}, the Neumann boundary condition is given as 
\begin{equation}
    \bb n \cdot \bb J = \rho_{\rm g} \sqrt{\frac{R_{\rm g}T}{2\pi \mu}}.
    \label{eq_neumann}
\end{equation}
Equation (\ref{eq_gas}), supplemented with the boundary condition (\ref{eq_neumann}) can be solved with the FEM analogously to the FEM solution of the energy balance equation (\ref{eq_diff}) by changing the coefficients and the right-hand side accordingly and introducing the boundary integral for the Neumann boundary condition (which is zero for the energy equation).
 
\begin{figure}[htb]
%\begin{center}
\includegraphics[width=0.49\textwidth]{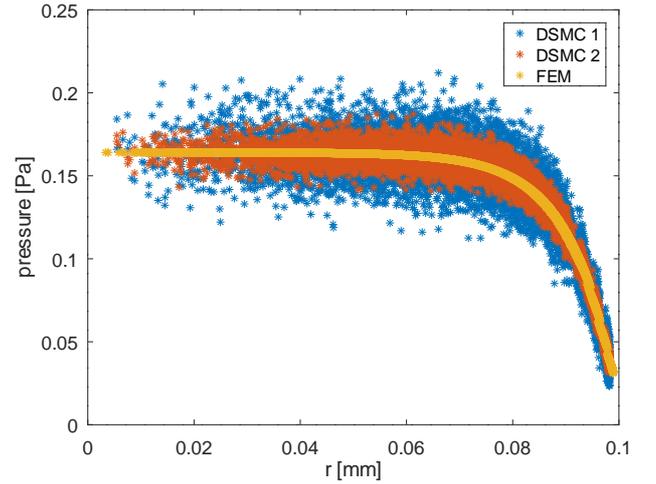}
\caption{Pressure as a function of distance from the center of the sphere calculated by the DSMC and FEM. The DSMC 2 solution has twice the number of the DSMC molecules than the DSMC 1 solution to demonstrate convergence.}
%\end{center}
\label{fig1}
\end{figure}

First, we considered the simplest possible case; an isothermal $T=\SI{200}{\kelvin}$ spherical particle with the radius $r=\SI{0.1}{\milli\metre}$ and  $S=\SI{10000}{\metre}^{-1}$ in a steady state. No other energy sources were included. We solved the problem using the DSMC and the FEM. Two different DSMC solutions were computed, namely, DSMC 1 and DSMC 2, the latter having twice the number of the DSMC molecules than the former. Figure \ref{fig1} shows the pressure sampled at the barycenter of each tetrahedron as a function of the distance from the center of the spherical particle. The solutions are in good agreement with each other but the DSMC solutions have more noise than the FEM solution. The DSMC noise decreases with the increasing number of the DSMC molecules, as expected. 

It is interesting to note that when $S$ is constant, Equation (\ref{eq_gas}) is reduced to the Helmholtz equation in a steady state with the wavenumber squared given by 
\begin{equation}
    k^2 = -\sqrt{\frac{R_{\rm g} T}{2\pi\mu}}S\Phi^{-1} D^{-1}.
\end{equation}
The fundamental solution, that is, the Green's function for the Helmholtz equation is written as
\begin{equation}
    G(\bb r, \bb r') = \frac{e^{ik|\bb r - \bb r'|}}{4\pi |\bb r - \bb r'|}
,\end{equation}
where $\bb r$ and $\bb r'$ are the source and observation points, respectively, and $i$ is the imaginary unit. Thus, with the imaginary $k$ the solution is an exponentially decaying evanescent wave, and the pressure drop near the surface can be very steep when $SD^{-1}$ is large. This also means that in order to solve Equation (\ref{eq_gas}) with the standard continuous Galerkin FEM, an extremely fine mesh is needed at the sublimation front, which can make computational time prohibitively long.

\begin{figure*}[ht]
\centering
\includegraphics[width=0.49\textwidth]{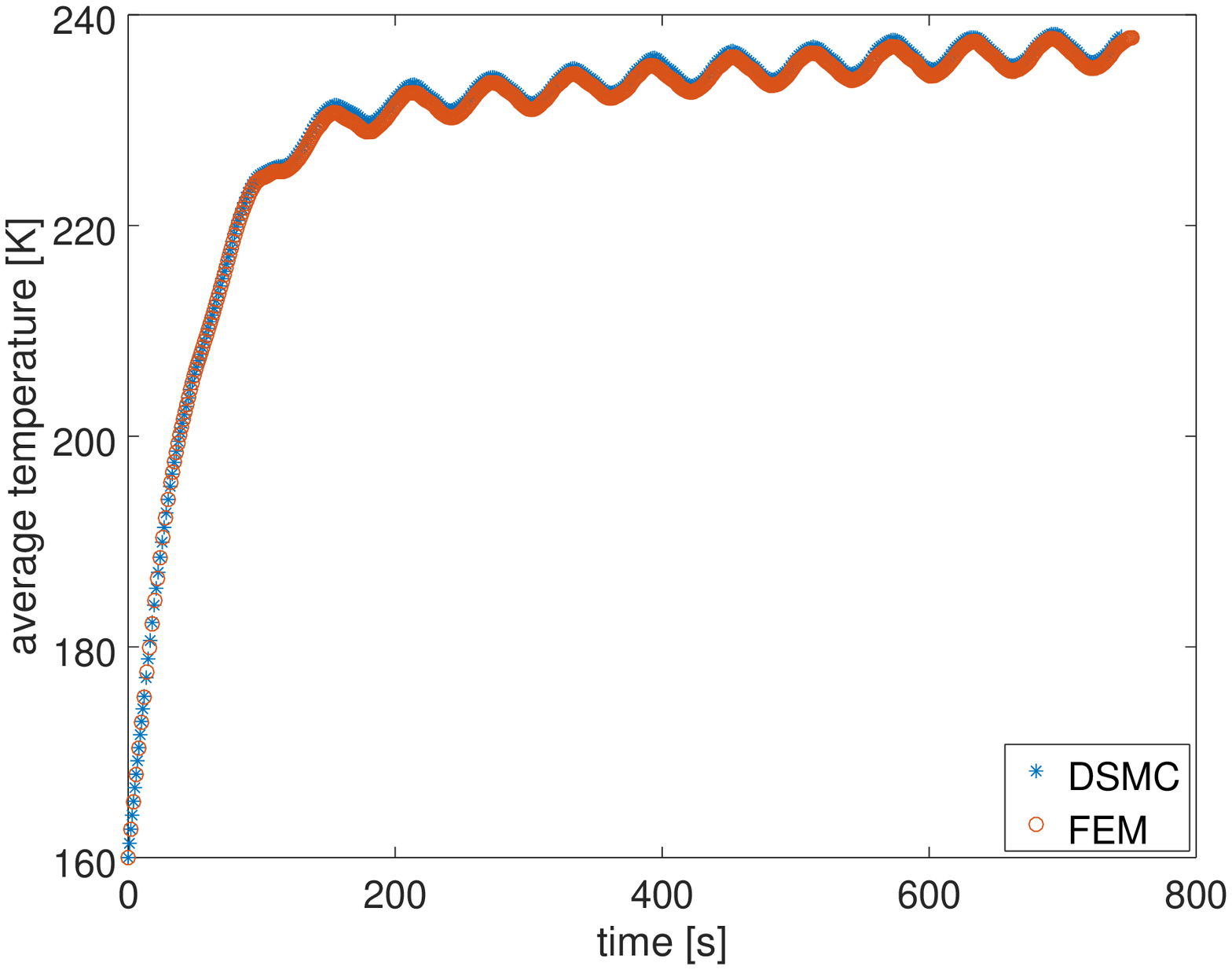}
\includegraphics[width=0.49\textwidth]{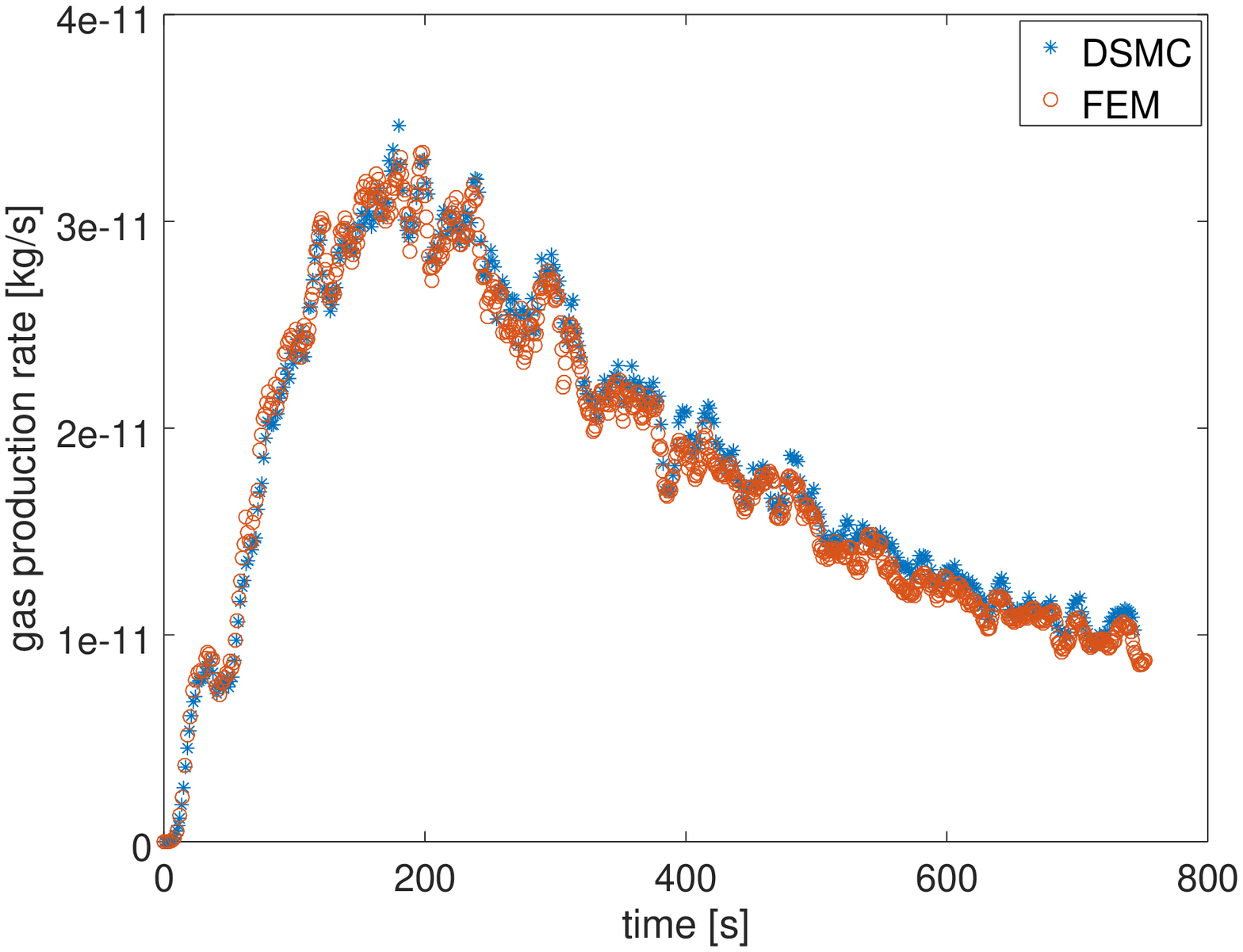}
\includegraphics[width=0.49\textwidth]{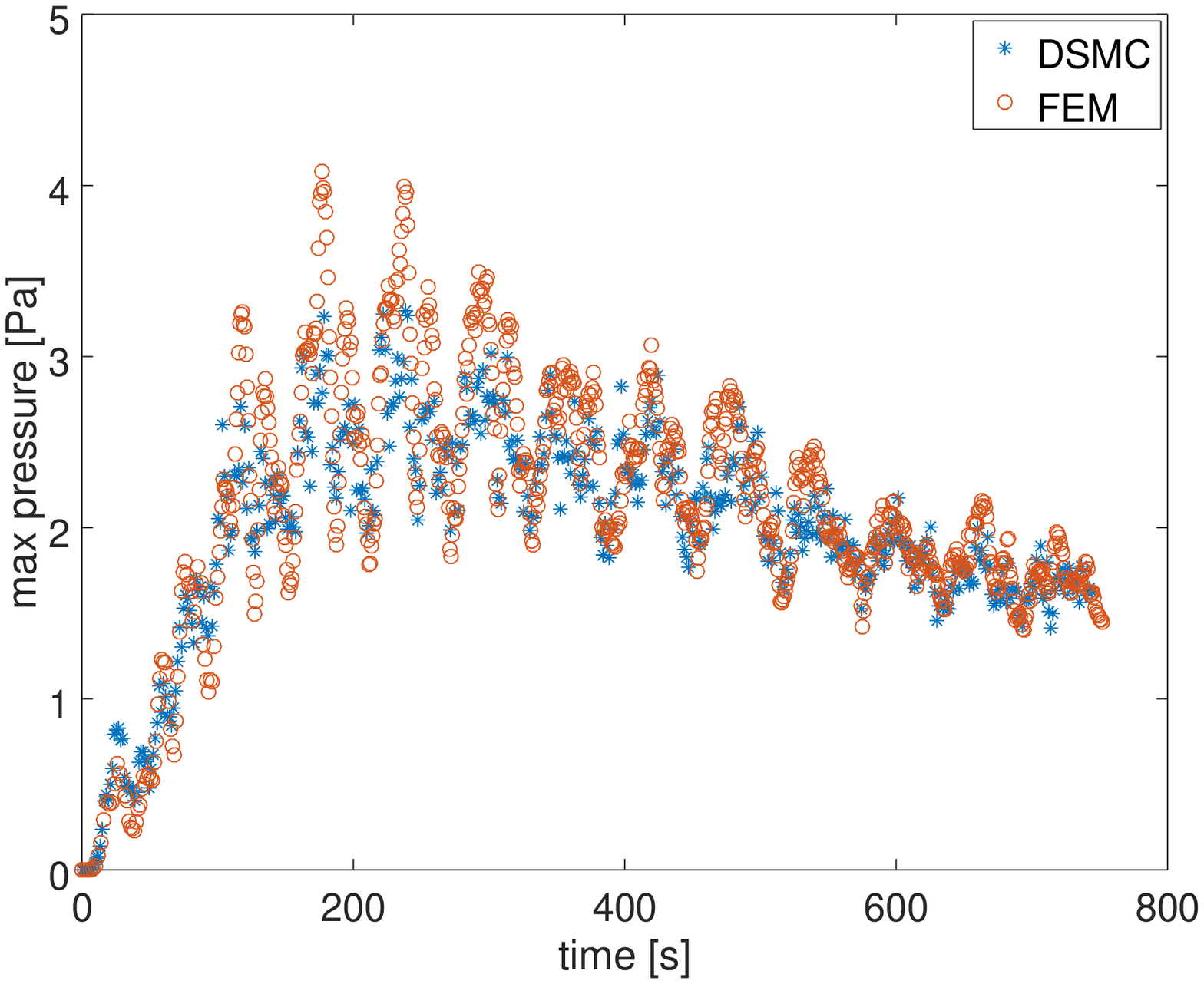}
\includegraphics[width=0.49\textwidth]{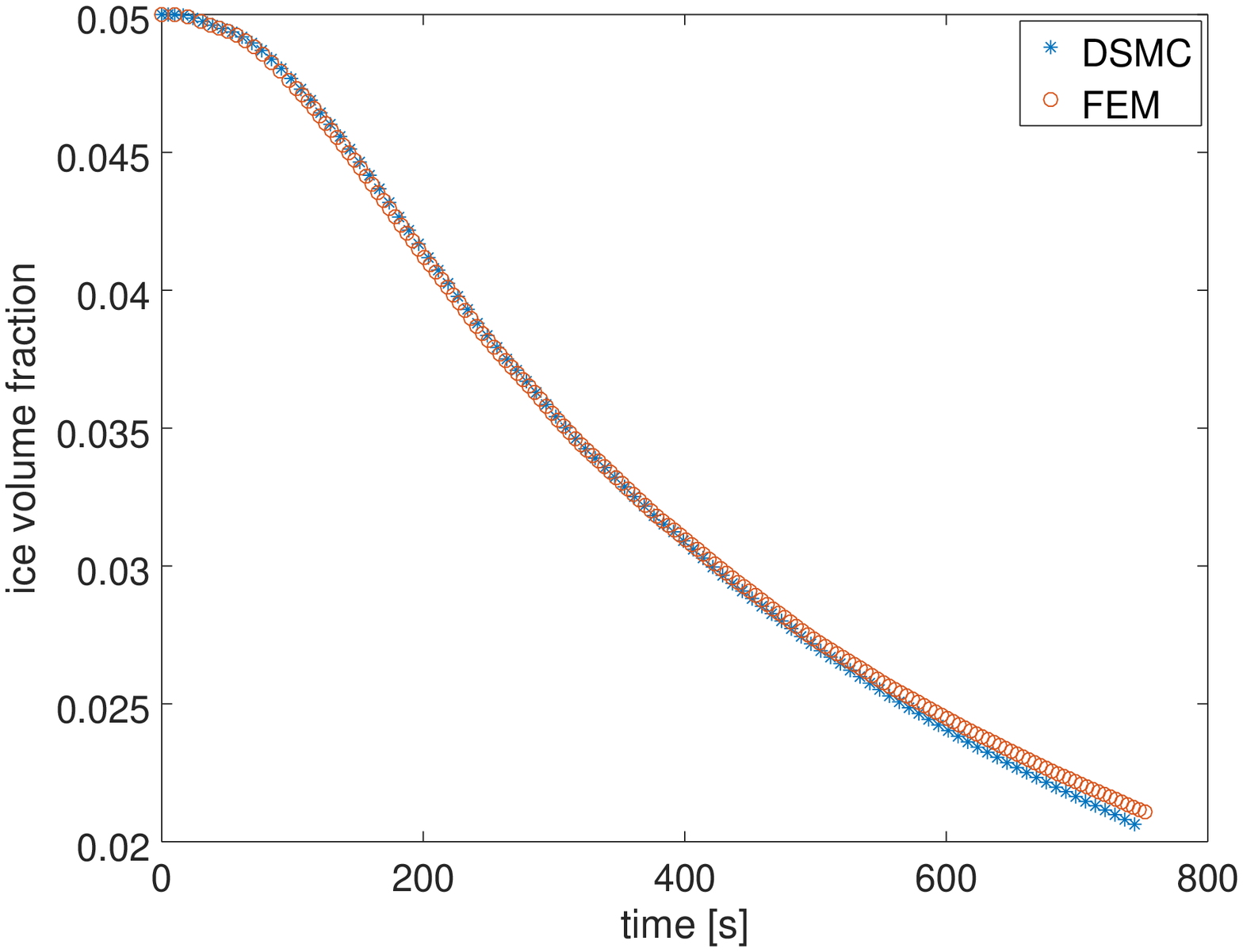}
\caption{Comparisons between thermal modelling results obtained by using the DSMC and FEM solutions for gas transport. The test particle has an irregular shape, with the volume equivalent radius of \SI{0.5}{\milli\metre}. The surface-to-volume ratio was set to S=\SI{1}{\metre^{-1}}.}
%\end{center}
\label{fig2}
\end{figure*}

Next, we simulated an irregularly-shaped dust particle under solar radiation at 1.35\,AU. The initial temperature of the particle was $\SI{160}{\kelvin}$ and the volumetric ice content was 0.05. The surface-to-volume ratio parameter was set to $S=\SI{1}{\metre}^{-1}$ to allow for an efficient FEM solution. The volume equivalent radius of the particle was $\SI{0.5}{\milli\metre}$ and it was spinning 1rpm with the spin axis pointing perpendicular to the direction of the Sun. The solar illumination was switched on at $t=\SI{0}{\second}$. The time evolution of the average temperature, gas production rate, maximum pressure averaged over a tetrahedron inside the particle and the ice volume fraction are plotted in Fig. \ref{fig2} computed by using the DSMC and FEM for the gas transport. We observed an excellent agreement between the two different methods. We note, however, that the boundary condition used in the FEM does not account for the gas flow back to the particle, whereas it is taken into account in the DSMC method. The gas production rate and maximum pressure peak at $t\approx \SI{200}{\second}$ and then they decrease. This happens because sublimation creates an insulating dry dust layer on the particle's surface that dampens the energy transport into the interior of the particle.

\subsection{Application to cometary dust}

Next, we studied thermal properties of dust in a cometary coma at 1.35\,AU. We assumed that the ejected dust particles have temperature $T_0 = \SI{160}{K}$ and contain five volume percent of water ice uniformly distributed into the micropores at $t=0$. Thus, the dust particle was assumed to originate below the hot surface layer in order to contain ice. Here, we do not account for the ejection mechanism as our goal is to study the thermal evolution of icy dust particles once they have been ejected from the nucleus and exposed to the direct sunlight. In the simulation, we assumed that the particles are rotating around the axis perpendicular to the solar direction with the angular speed of 1\,rpm. The results were not averaged over an ensemble of particles due to computational time restrictions.  

\begin{figure}[ht]
\includegraphics[width=0.49\textwidth]{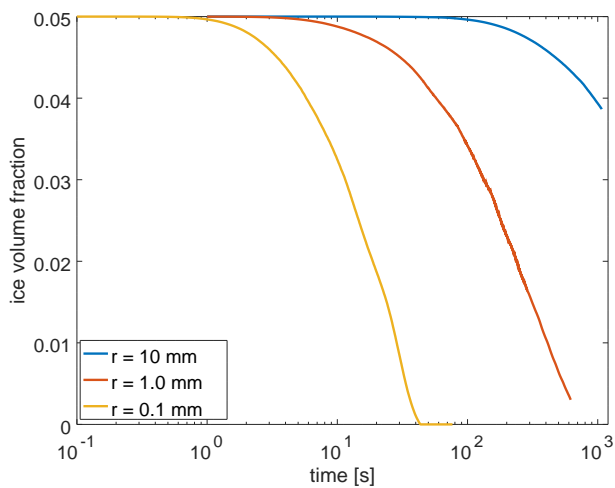}
\caption{Ice volume fraction as a function of time for different particle sizes.}
\label{fig3}
\end{figure}

Figure \ref{fig3} plots the ice volume fraction as a function of time for \SI{0.1}{\milli\metre}-, \SI{1}{\milli\metre}-, and \SI{10}{\milli\metre}-sized particles. We observed that the particles with $r = \SI{0.1}{\milli\metre}$ run out of ice in approximately a few tens of seconds after the ejection, $r = \SI{1}{\milli\metre}$ in ten minutes, and $r = \SI{10}{\milli\metre}$ particles can be extrapolated to stop subliming after a few hours. We expect that either they run out of ice or create an insulation surface layer, completely dumping sublimation. The time integration was stopped after 20 days of computing using 24 cores for the particle with $r = \SI{10}{\milli\metre}$. The smaller particles took less than a week each with 24 CPUs to complete the simulation. The biggest bottleneck in the simulations of large particles is the non-linearity of Equation (\ref{eq_diff}). A very small time step $\tau$ is required to solve the non-linear equation prohibiting long time evolution simulations. Thus, using a more sophisticated non-linear iterative solver or finding an efficient preconditioner would be an interesting topic for future research. Also, the DSMC simulation becomes slower with the increasing particle size as the diffusion time is proportional to $r^2$. Thus, for large particles, a continuous gas transport model would be preferred, assuming that the stability problem appearing for high $S$ due to the exponentially decaying pressure profiles can be solved. 

\begin{figure}[ht]
\includegraphics[width=0.49\textwidth]{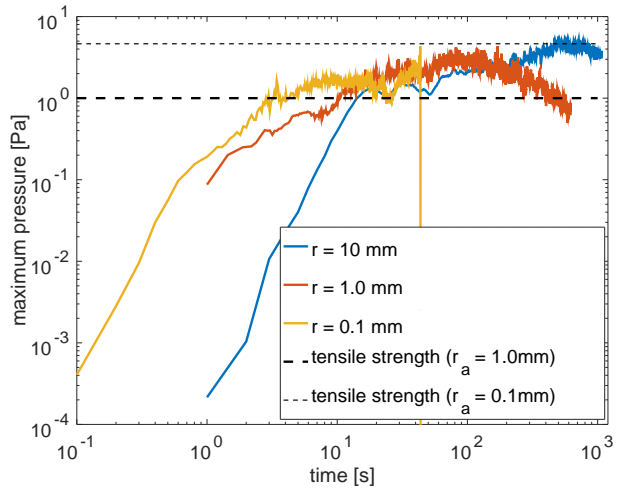}
\caption{Maximum pressure as a function of time for different particle sizes. }
\label{fig4}
\end{figure}

The sublimation of ices increases pressure inside the particles as demonstrated in Fig. \ref{fig4}. If the pressure is high enough, the particle may disintegrate. An experimentally verified model \citep{Skorov2012,Blum2014, Brisset2016} for the effective tensile strength of a particle made of aggregates is given by $T_{\rm eff} \approx (r_{\rm a} / \SI{1}{\milli\metre})^{-2/3}$ Pa where $r_{\rm a}$ is the radius of aggregates. In our particle model, these aggregates can be considered to correspond the Voronoi cells with $r_{\rm a} = r/10$. The tensile strengths for the particles made of aggregates with $r_{\rm a} = \SI{0.1}{\milli\metre}$ and \SI{1}{\milli\metre} are also presented in Fig. \ref{fig4} as dashed lines. The maximum pressure for particles with $r < \SI{1}{\milli\metre}$ is less than \SI{3}{\pascal} which is smaller than the corresponding tensile strength of the experimental particles. Thus, it is unlikely that a particle with $r < \SI{1}{\milli\metre}$ will disintegrate into smaller pieces because of pressure. For larger particles, the pressure can be higher. This happens as the sublimation creates a dry hot layer on the particle's surface which helps to build up pressure at the sublimation front. As shown in Fig. \ref{fig4}, for the particle with $r = \SI{10}{\milli\metre,}$ the maximum pressure can reach the tensile strength of a particle made of aggregates with $r_a = \SI{0.1}{\milli\metre}$. This indicates that the particles with $r >\SI{10}{\milli\metre}$ may start fragmenting into smaller pieces but we expect the disintegration to stop at mm-sized fragments. However, a detailed investigation of the fragmentation of particles is beyond the scope of this study and would require a detailed structural mechanical analysis. Albeit, we note that if such fragmentation would occur, it would have a significant effect on the sublimation rate as the insulating layer would be removed from time to time, allowing heat to directly access to the icy part.    

\begin{figure}[htb]
\includegraphics[width=0.49\textwidth]{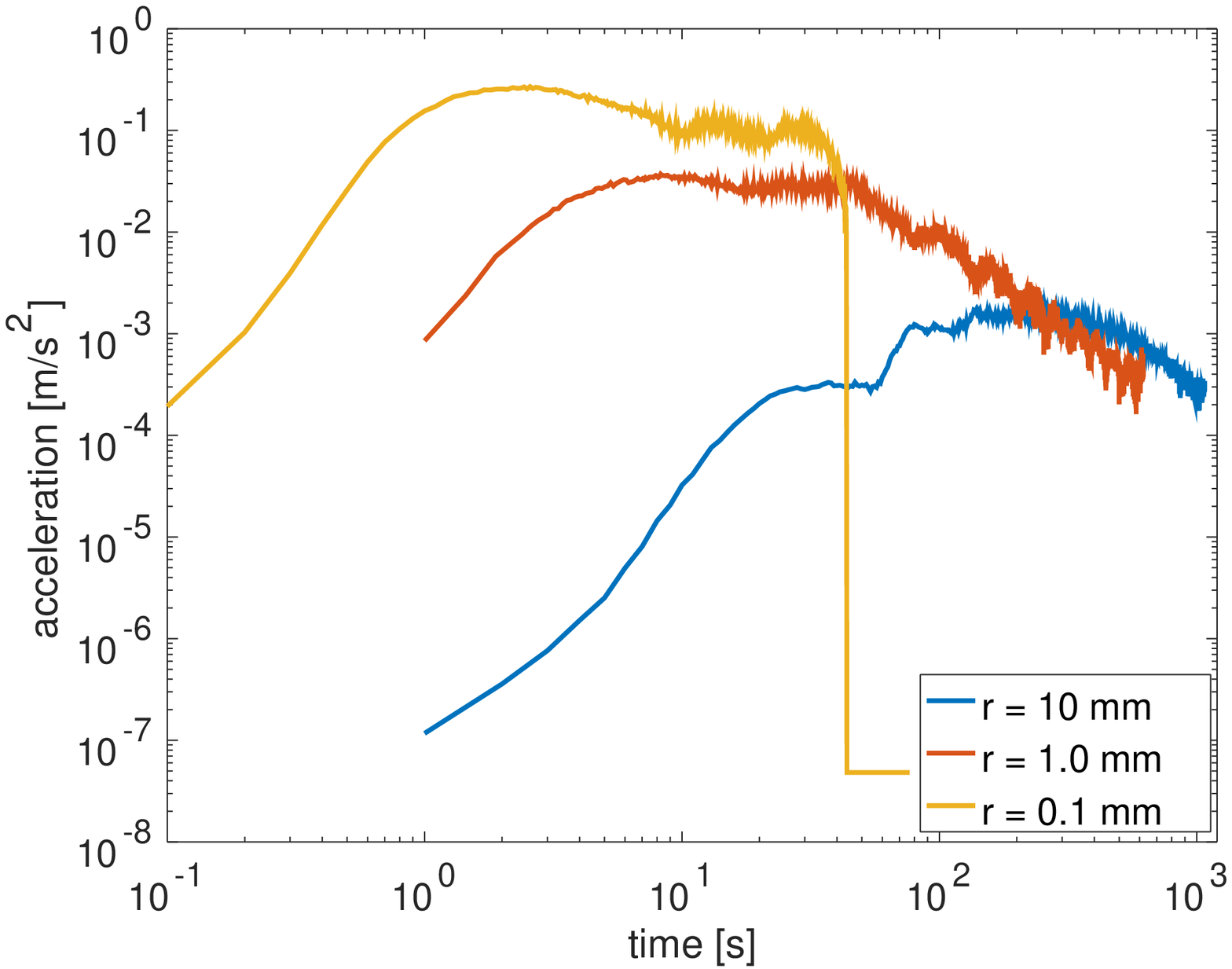}
\caption{Acceleration due to outgassing as a function of time for different particle sizes.}
\label{fig5}
\end{figure}

\begin{figure}[htb]
\includegraphics[width=0.49\textwidth]{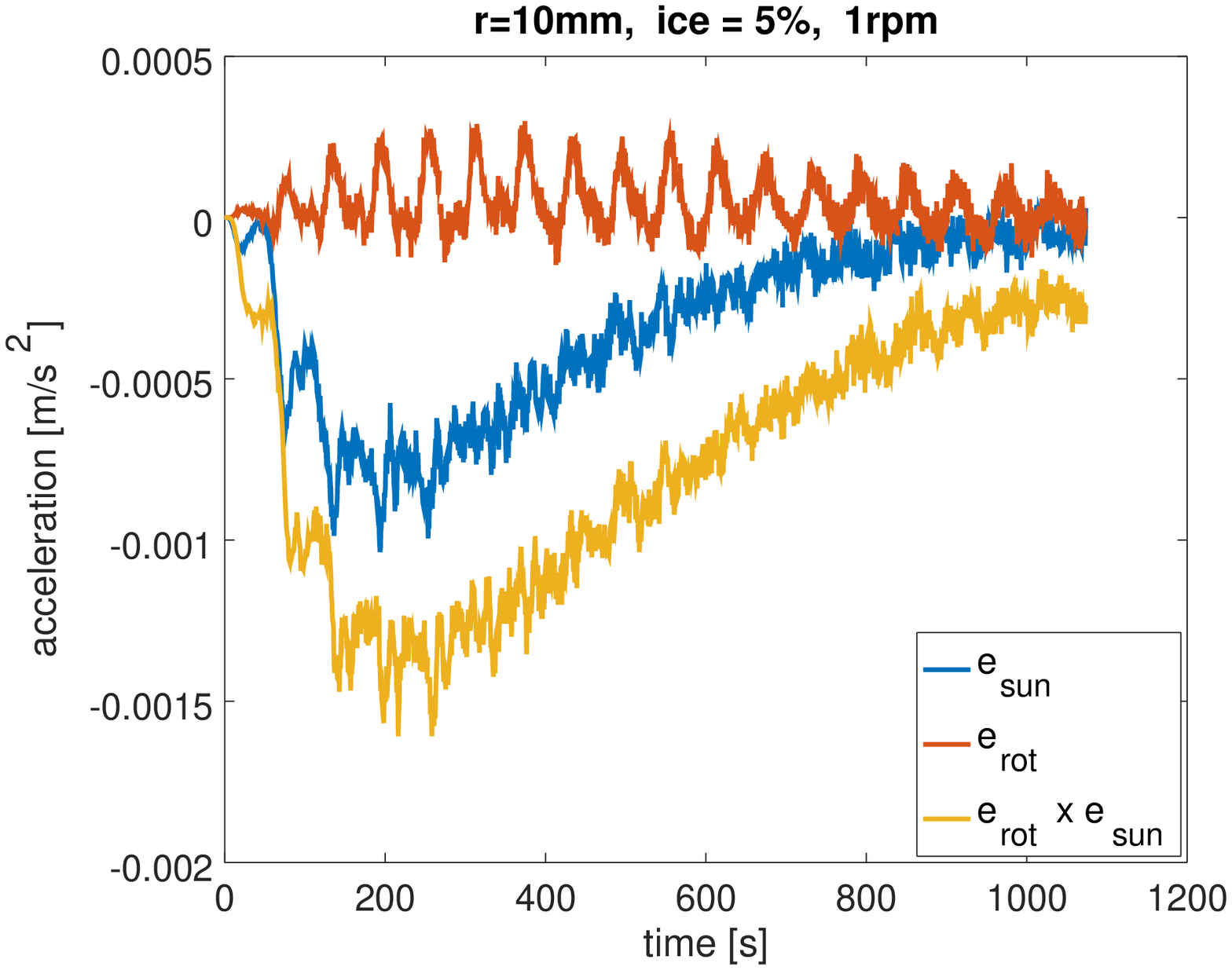}
\caption{Acceleration vector components for a particle with $r = 1$cm rotating 1 rpm about the axis, $\rm e_{\rm rot}$, perpendicular to the solar direction, $\rm e_{\rm sun}$, due to outgassing as a function of time. }
\label{fig6}
\end{figure}

Since the particles hold temperature gradients, outgassing of water vapour is anisotropic. This introduces a rocket force that accelerates the particles. The acceleration due to the rocket force is plotted in Fig. \ref{fig5}. The acceleration decreases with increasing particle size. This is clear as the sublimation rate is proportional to the illuminated area and the mass is proportional to the volume, the acceleration roughly scales linearly with the inverse of size. For small particles, the acceleration is high but it only occurs for a few seconds, whereas for large particles, the acceleration is small but it acts for a considerable time, which may have an effect on the particle's trajectory. The direction of the acceleration depends on the rotation state of the particle. For a slowly rotating particle, the force points towards the antisolar direction whereas for a fast rotating particle the direction is shifted towards the direction perpendicular to the rotation axis, $\rm e_{rot}$, and to the axis towards the Sun, $\rm e_{sun}$, as presented in Fig. \ref{fig6}.

Finally, we studied how the surface-to-volume ratio, $S,$ in Equation (\ref{eq_production_rate}) affects the results. Since sublimation requires that $P_{\rm sat} -P > 0$ and it depends on the surface-to-volume ratio of the ice surface, $S_{\rm ice},$ whereas condensation requires that $P_{\rm sat}-P < 0$ and it depends on the total surface-to-volume ratio $S_{\rm tot} = S_{\rm ice} + S_{\rm dust}$, we define
\begin{equation}
S= \left\{
\begin{array}{c}
    S_{\rm ice} \,\text{if}\, P_{\rm sat} - P > 0 \\
    S_{\rm tot} \,\text{if}\, P_{\rm sat} - P < 0.
    \end{array}
    \right.
\end{equation}
If ice fully covers the dust grains, it is clear that $S_{\rm ice} = S_{\rm tot}$. If the particle is an aggregate of ice and dust grains, then $S_{\rm ice} < S_{\rm tot}$. For equisized ice grains: 
\begin{equation}
S_{\rm ice} = 3v_{\rm ice}/r_{\rm ice}
\end{equation}
where $v_{\rm ice}$ is the volumetric filling factor of ice grains of radius $r_{\rm ice}$. Thus, $S$ depends on how ice is distributed inside the particle. We assumed that $v_{\rm ice} = 0.05$ and compared four different ice distributions. We also assumed that the ice distribution does not change the scattering, absorption and emission properties. In the first case, ice covered all the dust grains giving $S_{\rm ice} = \SI{1e7}{\metre^{-1}}$. In the second, the ice grains of radius $r_{\rm ice} = \SI{0.2}{\micro\metre}$ were evenly distributed inside the particle giving $S_{\rm ice} = \SI{7.5e5}{\metre^{-1}}$. In this case, the ice grains were approximately the same size as the dust grains. In the third, the ice grains were bigger, $r_{\rm ice} = \SI{2}{\micro\metre,}$ than the dust grains and $S_{\rm ice} = \SI{7.5e4}{\metre^{-1}}$. In the last case, the ice grains were much bigger than dust grains and $r_{\rm ice} = \SI{20}{\micro\metre}$.

\begin{figure}[htb]
\includegraphics[width=0.49\textwidth]{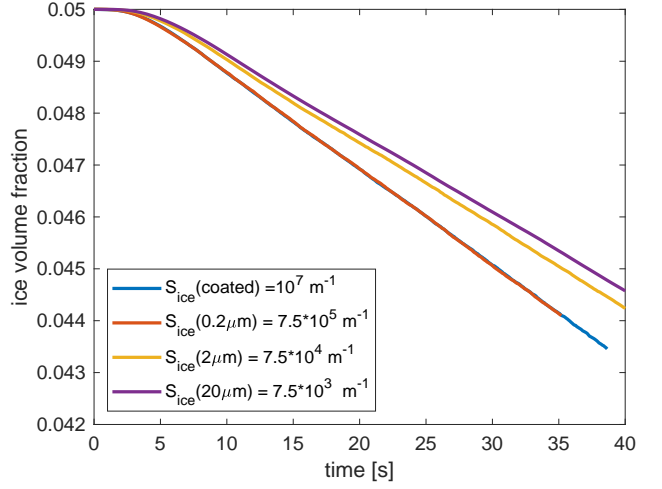}
\includegraphics[width=0.49\textwidth]{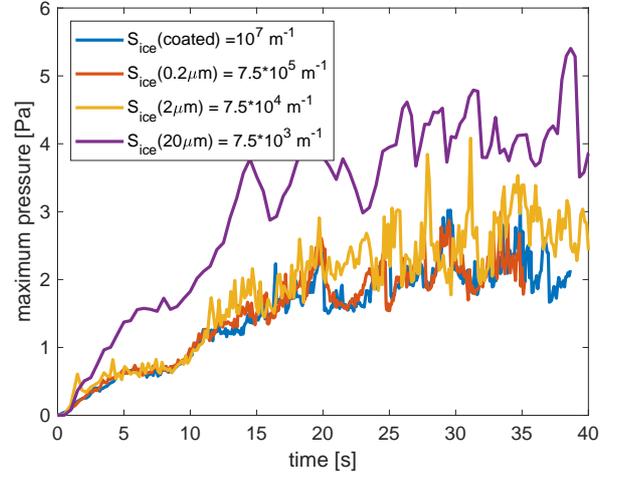}
\caption{Ice volume fraction and the maximum pressure as a function of time computed for varying ice surface-to-volume ratios, $S_{\rm ice}$, corresponding to fully coated dust grains, and ice grains of sizes \SI{0.2}{\micro\metre}, \SI{2}{\micro\metre,} and \SI{20}{\micro\metre} uniformly mixed with dust grains.}
\label{fig7}
\end{figure}

Comparisons for the total ice volume fraction and the maximum pressure as a function of time are presented in Fig. \ref{fig7}. We observed that when $S_{\rm ice}$ is large enough, the system is saturated and increasing $S_{\rm ice}$ does not affect the results. This is because gas diffusion gives the upper limit to the total sublimation rate. When $S_{\rm ice}$ is small enough, the total sublimation rate is smaller and the system is no longer saturated and limited by gas diffusion but is limited by the ice surface area. In such a case, the received energy is used more to increase the temperature of the particle which in turn allows higher local pressure. In fact, as seen from Equation (\ref{eq_gas}), the thickness of the sublimation front depends on the length scale, $h,$ and the diffusion coefficient, $D,$ and is proportional to the factor, $h S D^{-1}$. When the factor is large, the sublimation front is thin compared to the particle's size and the total sublimation rate depends on the macroscopic icy surface area rather than the microscopic surface-to-volume ratio, $S$. It is, thus, difficult to retrieve information on how the ice grains are mixed with the dust grains in the microscopic level from the total gas production rate.

\section{Conclusions}

We presented a novel thermophysical model and its numerical solution for icy cometary dust particles. The model employs the radiative transfer with reciprocal transactions for the radiative heat transport and the Boltzmann equation together with the Hertz-Knudsen formula for gas transport solved by the DSMC methods. Energy changes related to absorption, thermal emission, and phase changes were incorporated as source terms into the energy balance equation with the Fourier's heat conduction and solved by the FEM.    

The developed method allows for thermal analysis of up to cm-sized particles composed of submicrometer-sized grains mixed with ice with a moderate computing power. The method can find applications in understanding the thermal physics of cometary dust particles and explaining experimental laboratory measurements, and help us to develop more accurate approximate thermal modelling methods.  

We also showed that water ice sublimation inside large dust particles may generate enough pressure to reach the tensile strength and, thus, to possibly disintegrate the particles when comets are close to the Sun. Also, outgassing of water vapour can play a crucial role in the dynamics of icy dust particles after having been lifted off the comet's nucleus. Both of these processes should leave observable effects on the remote observables via changes in the size and spatial distributions. 

Finally, we showed that the thermal evolution of large dust particles is quite insensitive to the microscopic mixing scale of dust and ice, which effectively makes modelling easier. On the other hand, getting information on the mixing scale and ratio of dust and ice from the total gas production rate is not trivial.  

\begin{acknowledgements}
This work has been funded by the ERC Starting Grant No. 757390 Comet and Asteroid Re-Shaping through Activity (CAstRA). Computational resources have been provided by Gesellschaft f\"{u}r Wissenschaftliche Datenverarbeitung mbH G\"{o}ttingen (GWDG).
\end{acknowledgements}

   \bibliographystyle{aa} % style aa.bst
   \bibliography{refs.bib} % your references Yourfile.bib

\end{document}